# A Note on Linear Force Model in Car Accident Reconstruction

Milan Batista

University of Ljubljana, Faculty of Maritime Studies and Transportation

Pot pomorscakov 4, Slovenia, EU

milan.batista@fpp.edu

(Nov. 2005, updated)

**Abstract**

In the paper the linear force model used in car accident simulation programs is discussed. A model of restitution is proposed and the possible interpretation of CRASH coefficients is also discussed.

**1 Basic assumptions**

In a fixed barrier crush test (Figure 1), at the end of the compression phase the initial kinetic energy is absorbed to deformation. Thus one has from the conservation of energy

$$\frac{mv^2}{2} = E_m = \int_0^{\delta_m} F(\delta)d\delta \qquad (1)$$

where

- $F$    force
- $m$    mass
- $v$    initial velocity
- $W$    work (absorbed kinetic energy)
- $\delta$    crush
- $\delta_m$    maximal dynamics crush

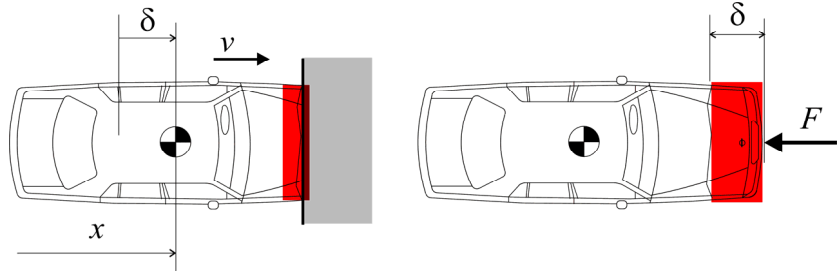

**Figure 1.** Fixed barrier crush test

If one assumes that force is, in the compression phase, proportional to crush i.e. if one use SMAC force model ([2])

$$F = K\delta \qquad (2)$$

where $K$ is stiffness, then (1) reduces to

$$\frac{mv^2}{2} = \frac{K\delta_m^2}{2} \qquad (3)$$

**Table 1.** Some values of initial $K$ from NHTSA NCAP tests ([7])

($R^2$ is correlation coefficient squer)

| Test No | Make | Model | Year | Mass [kg] | WB [mm] | Displ. [mm] | K [kN/m] | $R^2$ | $b_1 = \sqrt{K/m}$ [1/s] |
|---|---|---|---|---|---|---|---|---|---|
| 2319 | Audi | A4 | 1996 | 1763 | 2620 | 346 | 2319.8 | 0.95 | 36.3 |
| 4286 | Audi | A4 | 2002 | 1820 | 2645 | 291 | 2640.1 | 0.97 | 38.1 |
| 4248 | BMW | 325I | 2002 | 1731 | 2727 | 273 | 1496.8 | 0.97 | 29.4 |
| 4560 | BMW | X5 | 2003 | 2400 | 2820 | 320 | 1716.3 | 0.98 | 26.7 |
| 4444 | BMW | Z4 | 2003 | 1630 | 2500 | 250 | 1233.0 | 0.95 | 27.5 |
| 4491 | MB | C230 | 2003 | 1704 | 2713 | 299 | 1444.1 | 0.98 | 29.1 |
| 4266 | Toyota | Corolla | 2003 | 1350 | 2600 | 134-285 | 1595.7 | 0.95 | 34.4 |
| 3051 | VW | Beetle | 1999 | 1573 | 2527 | 293 | 1447.2 | 0.95 | 30.3 |
| 3239 | VW | Passat | 2000 | 1695 | 2715 | 329 | 2311.3 | 0.95 | 36.9 |

After the compression phase the restitution begins. Regardless of the mechanism of the restitution process one has at the end, if impact velocity is sufficiently high, residual crush. So the assumption is made that maximal dynamics crush can be split into two parts

$$\delta_m = \delta_0 + \delta_r \quad (4)$$

where $\delta_0$ is the recoverable part of crush and $\delta_r$ is residual crush. **Here the assumption is made that $\delta_0$ is constant; i.e., independent of residual crush.** In this context there is limited impact velocity $v_0$ where all the crush is recoverable. From conservation of energy one has the recoverable part of energy

$$E_0 = \frac{mv_0^2}{2} = \frac{K\delta_0^2}{2} \quad \Rightarrow \quad v_0 = \sqrt{\frac{K}{m}}\delta_0 \quad (5)$$

Thus, for velocities $v \geq v_0$ the recoverable part of energy is always the same. A model of force correspondent to the above assumption is well known ([3],[6]) and is shown on Figure 2. Note that in this model stiffness $K$ is the same for loading and unloading. The model in which the stiffness for loading and unloading phase of crush are different was developed by McHenry ([3],[4])

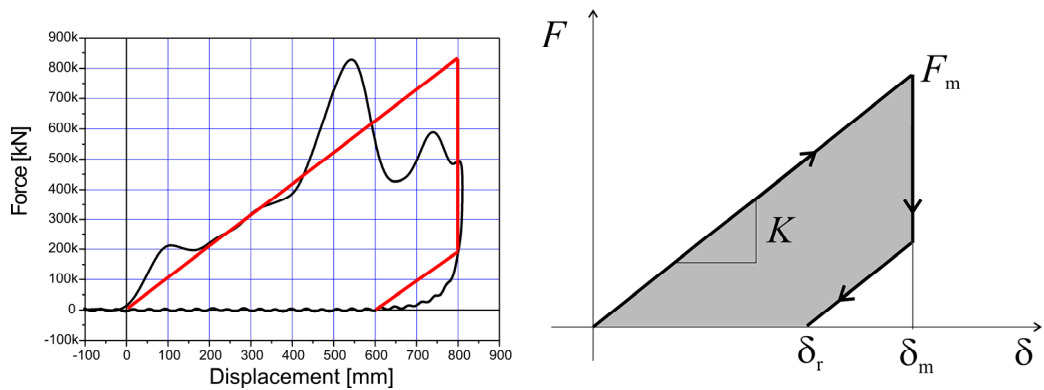

**Figure 2.** Force model

## 2 Relation between initial velocity and residual crush

Substituting (4) into (3) one finds for $v \geq v_0$

$$\frac{mv^2}{2} = \frac{K(\delta_0 + \delta_r)^2}{2} \quad \Rightarrow \quad v = \sqrt{\frac{K}{m}}(\delta_0 + \delta_r) \tag{6}$$

Thus impact velocity can be represented as the linear function of residual crush (Figure 3)

$$v = b_0 + b_1 \delta_r \tag{7}$$

where $b_0$ and $b_1$ are parameters. Note that relation (7) is the Campbell starting hypothesis ([1],[3]). If one assumes the validity of (7) then parameters $b_0$ and $b_1$ can be determined experimentally by fixed barrier impact tests. Once they are obtained, the parameters $K$ and $\delta_0$ are obtained - by comparing (7) and (6) - as

$$K = mb_1^2 \qquad \delta_0 = \frac{b_0}{b_1} \tag{8}$$

Note that assumption $v \geq v_0$ implies $b_0 = v_0$.

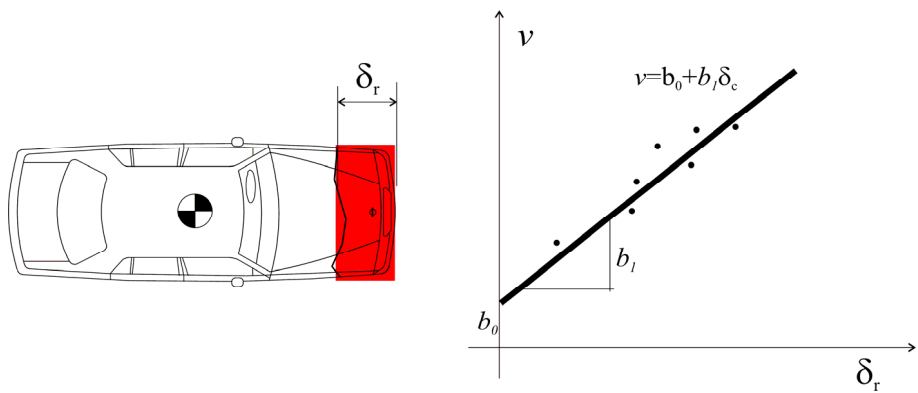

**Figure 3.** Linear connection between impact velocity and residual crush.

## 3 Relation between force and residual crush

The force at maximal crush is, from (2) and (4),

$$F = K\delta_m = K(\delta_0 + \delta_r) = K\delta_0 + K\delta_r \qquad (9)$$

By using (8) this can be written as ([1])

$$F = m(b_0 b_1 + b_1^2 \delta_r) \qquad (10)$$

Alternatively one can set

$$F = A + B\delta_r \qquad (11)$$

By comparing (10) and (11) and using (8) one finds connections between parameters

$$\boxed{A = K\delta_0 = mb_0 b_1 \qquad B = K = mb_1^2 \qquad \delta_0 = \frac{b_0}{b_1} = \frac{A}{B}} \qquad (12)$$

The maximum absorbed crush energy is, by using (11) and (4),

$$E_m = \frac{F\delta_m}{2} = \frac{A^2}{2B} + A\delta_r + \frac{B\delta_r^2}{2} = \frac{(A + B\delta_r)^2}{2B} \qquad (13)$$

This equation shows that maximum absorbed energy can be calculated if one knows residual crush. Once $E_m$ is known one can calculate the barrier impact velocity from (1)

$$v = \sqrt{\frac{2E_m}{m}} \qquad (14)$$

It is interesting to calculate the parameters in (12) from published CRASH coefficients recalculated to car class default width. The result is shown in Table 2. It seams that limit crush $\delta_0$ and consequently limit $v_0$ deformation is unrealistically high. This can be interpreted by the way the CRASH coefficients were calculated ([5]). Also by comparing $\sqrt{K/m}$ in Table 1 and 2 it is seen that in the first table the values are from 30 to 35, while in Table 2 for car class 1 to 3 it is about 20.

**Table 2.** Interpretation of CRASH coefficients A and B by (12)

|  |  | 1 | 2 | 3 | 4 | 5 | 6 |
|---|---|---|---|---|---|---|---|
| wheelbase | m | 2.055 | 2.408 | 2.581 | 2.804 | 2.985 | 2.769 |
|  |  | 2.408 | 2.581 | 2.804 | 2.985 | 3.129 | 3.302 |
| mass | kg | 1122 | 1251 | 1476 | 1794 | 2075 | 1818 |
| width | m | 1.544 | 1.707 | 1.844 | 1.956 | 2.027 | 2.007 |
| CRASH A | N/m | 52957 | 45417 | 55587 | 62426 | 56990 | 67161 |
| CRASH B | N/m$^2$ | 324475 | 296860 | 386608 | 234726 | 255437 | 869868 |
| $\delta_0$ | m | 0.163 | 0.153 | 0.144 | 0.266 | 0.223 | 0.077 |
| $K=B$ | kN/m | 501.0 | 506.7 | 712.9 | 459.1 | 413.3 | 149.4 |
| $b_1 = \sqrt{K/m}$ | 1/s | 21.13 | 20.13263 | 21.98 | 15.00 | 14.11 | 28.66 |
| $b_0 = v_0$ | m/s | 3.45 | 3.08 | 3.16 | 4.25 | 3.15 | 2.21 |
| $V_0$ | km/h | 12.4 | 11.1 | 11.4 | 15.3 | 11.3 | 7.0 |

**Note.** For the compact car class the default width is W=1.844 m, and the CRASH constants are A = 55.6 kN/m and B = 386.6 kN/m2. So for full width this becomes A*W=102.5 kN and B*W=712.9 kN/m. Then in the discussed model for full width one has K= B*W=712.9 kN/m. But for example for the compact car BMW 325I ( m= 1731 kg) the initial K for full width obtained from the NHTSA test 4248 is 1497 kN/m – more than twice as high. The rebound velocity in test was about 7.5 km/h so recoverable part of crush should be 0.071 m (which is in agreement with test). This give value (eq 12) A*W=106.2 kN which is comparable with the CRASH value. The average value of K (taking into account all the compression phase up to zero speed) for this case is about 810 kN/m. This value differs about 15% from default CRASH value. From average K one obtains b1=21.6 s$^{-1}$ and recoverable part of crush 0.096 m.

## 4 Restitution

By definition the restitution is for $\delta > \delta_0$ ([2]) by using (5) and (3)

$$e = \sqrt{\frac{E_0}{E_m}} = \sqrt{\frac{K\delta_0^2}{K\delta_m^2}} = \frac{\delta_0}{\delta_m} \qquad (15)$$

and when $\delta \leq \delta_0$ one has $e = 1$. This can be rewritten for the whole range of crush as (writing $\delta_m = \delta$)

$$e = \begin{cases} 1 & \delta \leq \delta_0 \\ \dfrac{\delta_0}{\delta} & \delta > \delta_0 \end{cases} \qquad (16)$$

Now if instead of crush the velocities are used in (15) for computation of energy one obtains

$$e = \sqrt{\frac{E_0}{E_m}} = \sqrt{\frac{mv_0^2}{mv^2}} = \frac{v_0}{v} \qquad (17)$$

or analogous to (16)

$$\boxed{e = \begin{cases} 1 & v \leq v_0 \\ \dfrac{v_0}{v} & v > v_0 \end{cases}} \qquad (18)$$

**This relation is interesting since it implies that upon non-elastic impact the rebound velocity is constant.** Indeed, by conservation of momentum and (18) one has

$$v \geq v_0 \ \wedge \ e = \frac{u}{v} = \frac{v_0}{v} \ \Rightarrow \ \boxed{u = v_0} \qquad (19)$$

where *u* is rebound velocity. That proposed interpretation is satisfied, it is seen, from figure 4, where the result of regression on test data is performed by using the form of restitution as (18).

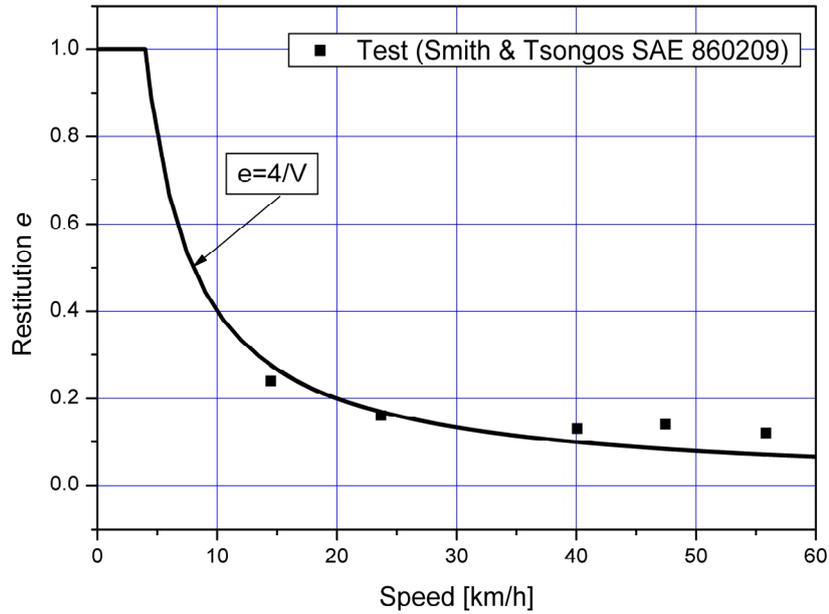

**Figure 4** Coefficient of restitution. Regression gives limit speed $v_0 = 4\,\text{km/h}$

Here it is interesting to note that the old SMAC program uses the following parameter called program variable ([2])

$$c = 1 - \frac{\delta_r}{\delta_m} = 1 - \frac{\delta_r}{\delta_0 + \delta_r} = \frac{\delta_0}{\delta_0 + \delta_r} = \frac{\delta_0}{\delta_m} \qquad (20)$$

By comparing (15) and (20) one finds that SMAC parameter *c* is equal to restitution coefficient *e*. In the quoted reference the connection between *c* and *e* is set to

$$c = 1 - \sqrt{1 - e^2} \qquad (21)$$

The origin of discrepancy is in the interpretation of returned energy $E_0$.

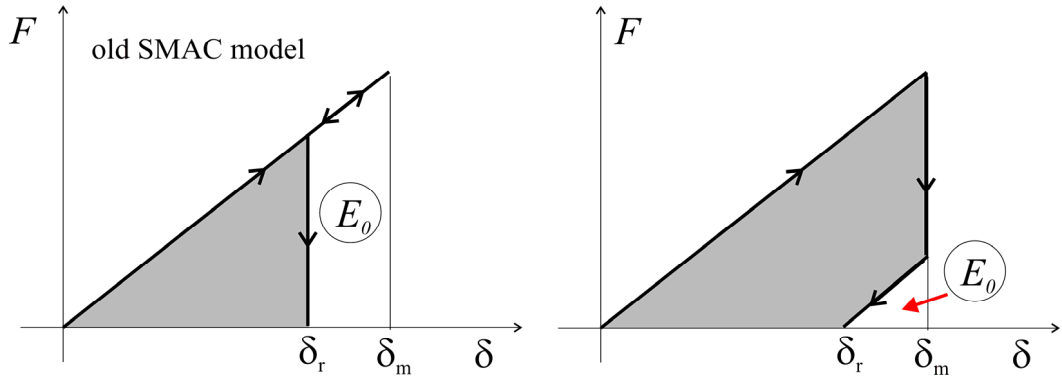

**Figure 5.** Comparison of models

By SMAC interpretation one has (Figure 5)

$$E_0 = \frac{K(\delta_m^2 - \delta_r^2)}{2} = \frac{K\delta_m^2}{2}\left(1 - \frac{\delta_r^2}{\delta_m^2}\right) = \frac{K\delta_m^2}{2}(2c - c^2) \qquad (22)$$

so

$$e = \sqrt{\frac{E_0}{E_m}} = \sqrt{2c - c^2} \qquad (23)$$

while for the present interpretation one has

$$E_0 = \frac{K(\delta_m - \delta_r)^2}{2} = \frac{K\delta_0^2}{2} \qquad (24)$$

which implies that $c = e$. The origin of this discrepancy is that old SMAC does not treat the rebound part of crush $\delta_0$ as constant while present model does.

**5 Conclusion**

In the present model the car is described by its mass $m$, stiffness $K$ and limit speed $v_0$ above the permanent crush. From the model one can interpret CRASH coefficients $A$ and $B$ similarly to what was done by Tamny ([6]). The restitution model is simple: the rebound velocity is constant and equals $v_0$. This is the consequence of the assumption that returned energy is independent of residual crush; i.e., it is constant.

## References

[1] K.L.Campbell. Energy Basis for Collision Severity. SAE 740565

[2] R.R.McHenry, Mathematical Reconstruction of Highway Accidents. PB-240 867, NTIS, 1975

[3] R.R.McHenry, B.G.McHenry, A Revised Damage Analysis Procedure for the CRASH Computer Program, SAE 861894

[4] R.R.McHenry, B.G.McHenry, Effects of Restitution in the Application of Crush Coefficients, SAE 970960

[5] R.A.Smith, J.T.Noga, Accuracy and sensitivity of CRASH, SAE 821169

[6] S.Tamny, The Linear Elastic-Plastic Vehicle Collision, SAE 921073

[7] http://www-nrd.nhtsa.dot.gov/